\documentclass[aps,twocolumn,showpacs,preprintnumbers,amsmath,amssymb]{revtex4}

\usepackage{graphicx}
\usepackage{dcolumn}
\usepackage{bm}

\begin{document}

\preprint{APS/1: Thin-shell-wormhole}

\title{Thin-shell wormholes from charged black holes in
 generalized dilaton-axion gravity}

\author{A. A. Usmani}
\email{anisul@iucaa.ernet.in} \affiliation{Department of Physics,
Aligarh Muslim University, Aligarh 202 002, Uttar Pradesh, India.}

\author{F. Rahaman}
 \email{farook\_rahaman@yahoo.com}
\affiliation{Department of Mathematics, Jadavpur University,
Kolkata 700 032, West Bengal, India}

\author{Saibal Ray}
\email{saibal@iucaa.ernet.in} \affiliation{Department of Physics,
Government College of Engineering \& Ceramic Technology, Kolkata
700 010, West Bengal, India}

\author{Sk. A. Rakib}
\email{farook\_rahaman@yahoo.com} \affiliation{Department of
Mathematics, Jadavpur University, Kolkata 700 032, West Bengal,
India}

\author{Z. Hasan}
\email{anisul@iucaa.ernet.in} \affiliation{Department of Physics,
Aligarh Muslim University, Aligarh 202 002, Uttar Pradesh, India.}

\author{Peter K. F. Kuhfittig }
\email{kuhfitti@msoe.edu} \affiliation{Department of Mathematics,
Milwaukee School of Engineering, Milwaukee, Wisconsin 53202-3109,
USA}

\date{\today}

\begin{abstract}\noindent
This paper discusses a new type of thin-shell wormhole
constructed by applying the cut-and-paste technique to
two copies of a charged black hole in generalized
dilaton-axion gravity, which was inspired by low-energy
string theory.  After analyzing various aspects of this
thin-shell wormhole, we discuss its stability to
linearized spherically symmetric perturbations.
\end{abstract}


\maketitle

\section{Introduction}
\noindent
The study of traversable wormholes has received considerable
attention from researchers for the past two decades.  Although
lacking observational evidence, wormholes are just as good a
prediction of the general theory of relativity as black holes.
In particular, we refer to the pioneering work of Visser
\cite{Visser1989}, who proposed a theoretical method for
constructing a new class of traversable Lorentzian wormholes
from black-hole spacetimes. This construction proceeds by
surgically grafting two Schwarzschild spacetimes
together in such a way that no event horizon is permitted to
form. The resulting structure is a wormhole spacetime in which
the throat is a three-dimensional thin shell. In recent years,
Visser's approach was adopted by various authors for constructing
thin-shell wormholes by similar methods, generally requiring
spherical symmetry
\cite{Poisson1995,Lobo2003,Lobo2004,Eiroa2004a,Eiroa2004b,Eiroa2005,
Thibeault2005,Lobo2005,Rahaman2006,
Eiroa2007,Rahaman2007a,Rahaman2007b,Rahaman2007c,
Lemos2007,Richarte2008,Rahaman2008a,
Rahaman2008b,Eiroa2008a,Eiroa2008b}. The approach is of
special interest because it minimizes the amount of exotic matter
required.  All the exotic matter is confined to the shell.

More recently, Sur, Das, and SenGupta \cite{Sur2005} discovered
a new black-hole solution for Einstein-Maxwell scalar field
systems inspired by low-energy string theory. They considered
a generalized action in which two scalar fields are minimally
coupled to an Einstein-Hilbert-Maxwell field in four
dimensions,
\begin{equation}
I = \frac{1}{2 \kappa} \int  d^4x \sqrt{-g}\left[ R -
\frac{1}{2}\partial_\mu\varphi \partial^\mu\varphi - W \right],
\end{equation}
where
\begin{multline*}
 W=\frac{1}{2}\omega(\varphi) \partial_\mu \zeta
 \partial^\mu\zeta - \alpha (\varphi,\zeta)
    F_{\mu\nu}F^{\mu\nu}\\
  - \beta (\varphi,\zeta)F_{\mu\nu} F^{\mu\nu\ast},
\end{multline*}
$\kappa = 8 \pi G $, $R$ is the curvature scalar, $F_{\mu\nu}$
is the Maxwell field tensor, while $\varphi$ and $\zeta$ are
two massless scalar or pseudo scalar fields, which are
coupled to the Maxwell field.  This coupling is described by
the functions $\alpha $ and $\beta$.  Here $\zeta $ acquires a
non-minimal kinetic term $\omega$. In the context of low-energy
string theory, fields $\phi$ and $\xi$ can be identified as
massless scalar dilaton and pseudo scalar axion fields,
respectively.

With the above action, Eq. (1), Sur, \emph{et al.}, \cite{Sur2005}
found the most general class of black-hole solutions and
obtained two types of metrics, classified as asymptotically flat
and asymptotically non-flat. Since we are interested in
obtaining a thin-shell wormhole from this new black hole, we
adopt the asymptotically flat metric given by
\begin{equation}\label{E:line1}
ds^2 = -f(r) dt^2 + f(r)^{-1}dr^2 + h(r) (d\theta^2+\sin^2\theta
d\phi^2),
\end{equation}
where
\begin{equation}
f(r) = \frac{(r-r_-)(r-r_+)}{(r-r_0)^{2-2n}(r+r_0)^{2n}},
\end{equation}
and
\begin{equation}
h(r) = \frac{(r+r_0)^{2n}}{(r-r_0)^{2n-2}},
\end{equation}
where, according to Ref. \cite{Sur2005}, the exponent $n$
is a dimensionless constant stricly greater than 0 and
stricly less than 1.  In addition, various other parameters
are given by
\begin{eqnarray}
r_{\pm} &=& m_0 \pm \sqrt{m_0^2 +r_0^2 -\frac{1}{8}\left(
\frac{K_1}{n} + \frac{K_2}{1-n}\right)},\\ r_0 &=&\frac{1}{16m_0}
\left(\frac{K_1}{n} - \frac{K_2}{1-n}\right),\\ m_0 &=& m -
(2n-1)r_0,\\ K_1 &=& 4n[ 4r_0^2 + 2r_0(r_+ +r_-) + r_+r_-],\\ K_2
&=& 4(1-n)r_+r_-, \ \ 0<n<1,\\ m &=& \frac{1}{16r_0} \left(
\frac{K_1}{n} - \frac{K_2}{1-n}\right) + (2n-1)r_0,
\end{eqnarray}
where $m$ is the mass of the black hole.  The parameters
$r_+$ and $r_-$ are the inner and outer event horizons,
respectively.  Also, $r=r_0$ is a curvature singularity;
the parameters obey the condition $r_0<r_-<r_+$.

In this paper we present a new kind of thin-shell wormhole
by surgically grafting two charged black holes in generalized
dilaton-axion gravity. The exotic matter required for its
physical existence may possibly be collected from scalar
fields that built the black holes.  Various aspects of this
thin-shell wormhole are analyzed, particularly the equation
of state relating pressure and density. Also discussed is
the attractive or repulsive nature of the wormhole, as well
as the energy conditions on the shell.  Our final topic is
a stability analysis to determine the conditions under which
the wormhole is stable to linearized radial perturbations.
A comparison to the stability of other thin-shell
wormholes in the literature is also made.

\section{Thin-shell wormhole construction}
\noindent
The mathematical construction of our thin-shell wormhole begins
by taking two copies of the black hole and removing from each
the four-dimensional region
\[
  \Omega^\pm = \{r\leq a \mid a>r_+\}.
\]
We now identify (in the sense of topology) the timelike
hypersurfaces
\[
  \partial\Omega^\pm = \{r=a \mid a>r_+\},
\]
denoted by $\Sigma$.  The resulting manifold is geodesically
complete and consists of two asymptotically flat regions
connected by a throat. The induced metric on $\Sigma$ is
given by
\begin{equation}
               ds^2 =  - d\tau^2 + a(\tau)^2( d\theta^2 +
               \sin^2\theta d\phi^2),
\end{equation}
where $\tau$ is the proper time on the junction surface.  Using
the Lanczos equations
\cite{Visser1989,Poisson1995,Lobo2003,Lobo2004,Eiroa2004a,Eiroa2004b,Eiroa2005,
Thibeault2005,Lobo2005,Rahaman2006,
Eiroa2007,Rahaman2007a,Rahaman2007b,Rahaman2007c,Lemos2007,Richarte2008,Rahaman2008a,
Rahaman2008b,Eiroa2008a,Eiroa2008b}, one can obtain the surface
stress energy tensor $ S_{\phantom{i}j}^i=\text{diag}(-\sigma,p_{\theta},
p_{\phi})$, where $\sigma$ is the surface energy density and
$p_{\theta}$ and $p_{\phi}$ are the surface pressures.
The Lanczos equations now yield \cite{Eiroa2005}
\begin{equation}\label{E:sigma1}
\sigma = - \frac{1}{4\pi }\frac{h^\prime(a)}{h(a)}\sqrt{f(a) +
\dot{a}^2}
\end{equation}
and
\begin{equation}\label{E:pressure1}
p_{\theta} = p_{\phi} = p =  \frac{1}{8\pi
}\frac{h^\prime(a)}{h(a)}\sqrt{f(a) + \dot{a}^2} + \frac{1}{8\pi
}\frac{2\ddot{a} + f^\prime(a) }{\sqrt{f(a) + \dot{a}^2}}.
\end{equation}
To understand the dynamics of the wormhole, we assume  the radius
of the throat to be a function of proper time, or $ a =
a(\tau)$. Also, overdot and prime denote, respectively, the
derivatives with respect to $\tau$ and $a$. For a static
configuration of radius $a$, we obtain the respective values
of the surface energy density and the surface pressures. For
a static configeration of radius $a$, we obtain (assuming
$\dot{a} = 0 $ and $\ddot{a}= 0 $) from Eqs. (\ref{E:sigma1})
and (\ref{E:pressure1}),
\begin{equation}\label{E:sigma2}
\sigma = -
\frac{4[a+(1-2n)r_0]}{D}\frac{(a-r_-)(a-r_+)}{(a-r_0)(a+r_0)}
\end{equation}
and
\begin{equation}\label{E:pressure2}
p_{\theta}= p_{\phi} =p= \frac{2a-r_- -r_+}{D}
\end{equation}
where
\begin{equation}\label{E:D}
D=8\pi (a-r_0)^{1-n}(a+r_0)^{n} \sqrt{(a-r_-)(a-r_+)}.
\end{equation}
Observe that the energy density $\sigma$ is negative. The
pressure $p$ may be positive, however.  This would depend on
the position of the throat and hence on the physical parameters
$r_0$, $r_-$, and $r_+$ defining the wormhole. Similarly, $p +
\sigma$, $ \sigma + 2p  $  $ $, and  $\sigma+3p $, obtained by
using the above equations, may also be positive under certain
conditions, in which case the strong energy condition is
satisfied.

\begin{figure}
\begin{center}
\vspace{0.5cm}
\includegraphics[width=0.5\textwidth]{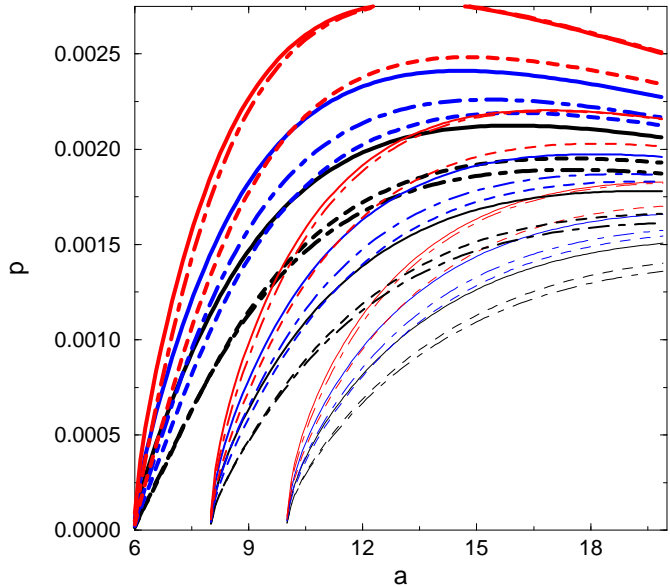}
\caption{Plot for $p$ versus $a$. The black, blue, and red colors
represent $n=0.98$, 0.5 and 0.02, respectively. For every color,
thin, thick, and thicker curves, respectively, represent $r_+=10$,
$8$, and $6$. For every combination of $r_+$ and $n$, we plot three
different sets, ($r_-=5$, $r_0=2$), ($r_-=5$, $r_0=3$), and
($r_-=4$, $r_0=2$), which are represented by chain and solid
curves, respectively. } \label{fig1}
\end{center}
\end{figure}

Keeping in mind the condition $r_+>r_->r_0$ for
different radii defining the wormhole, we plot $p$ versus $a$ in
Fig.~\ref{fig1}. We choose typical wormholes whose radii ($r_0$,
$r_-$, and $r_+$) fall within the range $2$ to $12$ kms.
Also taken into account is the sensitivity of the plots  with
respect to $n$, as described in the caption of the figure.

\section{The gravitational field}
\noindent
We now turn our attention to the attractive or repulsive
nature of our wormhole.  To perform the analysis, we
calculate the observer's four-acceleration
$a^\mu = u^\mu_{\,\,;\nu} u^\nu$, where
$u^{\nu} =  d x^{\nu}/d {\tau}
=(1/\sqrt{f(r)}, 0,0,0)$.  In view of the line
element, Eq. (\ref{E:line1}), the only non-zero component
is given by

\begin{equation}\label{E:acceleration}
a^r = \Gamma^r_{tt}
\left(\frac{dt}{d\tau}\right)^2 = \frac{1}{2}
\frac{Ar^2-Br+C}{(r-r_0)^{3-2n}(r+r_0)^{2n+1}} \end{equation}
where,
\[A= r_- +r_+ + 4nr_0 -2r_0,\]
\[ B= 2r_0^2 +(r_- + r_+)(4nr_0-2r_0)+ 2r_- r_+, \]
and
\[ C= r_0^2 (r_- + r_+)+(4nr_0-2r_0) r_- r_+ .\]

A radially moving test particle initially at rest obeys
the equation of motion
\begin{equation}\label{E:motion}
\frac{d^2r}{d\tau^2}= -\Gamma^r_{tt}\left(\frac{dt}{d\tau}
\right)^2 =-a^r.
\end{equation}
If $a^r=0$, we obtain the geodesic equation.  Moreover, a
wormhole is attractive if $a^r>0$ and repulsive if $a^r<0$.
These characteristics depend on the parameters $r_0$, $r_-$,
$r_+$, and $n$, the conditions on which can be conveniently
expressed in terms of the coefficients $A$, $B$, and $C$.
To avoid negative values for $r$, let us consider only the root
$r=(B+\sqrt{B^2-4AC})/(2A)$ of the quadratic equation
$Ar^2-Br+C=0$.  It now follows from Eq.~(\ref{E:acceleration})
that $a^r=0$ whenever
 \[ \left( r -\frac{B}{2A} \right)^2
    =  \frac{ B^2 -4AC}{4A^2} .\]
For the attractive case, $a^r>0$, the condition becomes
\[ \left(r -\frac{B}{2A} \right)^2
    >\frac{ B^2 -4AC}{4A^2}. \]
For the repulsive case, $a^r<0$, the sense of the inequality
is reversed.

\section{The total amount of exotic matter}
\noindent
In this section we determine the total amount of exotic matter
for the thin-shell wormhole.  This total can be quantified by the
integral
 \cite{Eiroa2005,Thibeault2005,Lobo2005,Rahaman2006,
Eiroa2007,Rahaman2007a,Rahaman2007b}
\begin{equation}
   \Omega_{\sigma}=\int [\rho+p]\sqrt{-g}d^3x.
\end{equation}
By introducing the radial coordinate $R=r-a$, w get
\[
 \Omega_{\sigma}=\int^{2\pi}_0\int^{\pi}_0\int^{\infty}_{-\infty}
     [\rho+p]\sqrt{-g}\,dR\,d\theta\,d\phi.
\]
Since the shell is infinitely thin, it does not exert any radial
pressure.  Moreover, $\rho=\delta(R)\sigma(a)$.  So
\begin{multline}\label{E:amount}
 \Omega_{\sigma}=\int^{2\pi}_0\int^{\pi}_0\left.[\rho\sqrt{-g}]
   \right|_{r=a}d\theta\,d\phi=4\pi h(a)\sigma(a)\\
=-\frac{16\pi[a+(1-2n)r_0]}{D}\left[\frac{(a-r_-)(a-r_+)}
{(a-r_0)^{2n-1}(a+r_0)^{1-2n}}\right].
\end{multline}
Here $D$ is given in Eq. (\ref{E:D}).

This NEC violating matter can be reduced by taking the
value of $a$ closer to $r_+$, the location of the outer
event horizon.  The closer $a$ is to $r_+$, however, the
closer the wormhole is to a black hole: incoming microwave
background radiation would get blueshifted to an extremely
high temperature \cite{tR93}.  On the other hand, it follows
from Eq. (\ref{E:amount}) that for $a\gg r_+$,
$\Omega_{\sigma}$ will  depend linearly on $a$: \\
\begin{equation}
\Omega_{\sigma} \approx -2a. \end{equation}

\section{An equation of state}
\noindent
Taking the form of the equation of state (EoS) to be
$p=w\sigma$, we obtain from Eqs. (\ref{E:sigma2}) and
(\ref{E:pressure2}),
\begin{equation}\label{E:EoS}
\frac{p}{\sigma}  = w = \frac{1}{4}  \frac{(2a-r_- -
r_+)(r_0^2-a^2)}{(a-r_-)(a-r_+)[a+(1-2n)r_0]}.
\end{equation}
Observe that if the location of the wormhole throat is
very large, i.e., if $a\rightarrow +\infty$, then
$w\rightarrow -\frac{1}{2}$.  On the other hand, if
$a\rightarrow r_{+}$ (from the right),
then $\omega\rightarrow -\infty$.
So the distribution of matter in the shell is of the
dark-energy type.  Now, purely mathematicall speaking,
if $a \rightarrow \frac{1}{2}(r_-+ r_+)$, then
$p \rightarrow 0$.  Since $\frac{1}{2}(r_-+ r_+)<r_+$,
however, such a dust shell is never found.

Our spacetime metric implies that the surface mass of this
thin shell is given by $M_{shell} =
4 \pi h(a) \sigma$. (For a static solution, we have
$\dot{a} = 0$ and $\ddot{a}= 0$.) Thus
\begin{multline}
M_{shell} =  2   \frac{[a+(1-2n)r_0]}{(r_0^2-a^2)}\times\\
(a-r_0)^{1-n}(a+r_0)^n\sqrt{(a-r_-)(a-r_+)} .
\end{multline}

Now observe that for $n=\frac{1}{2}$, the mass of the black
hole in Eq. (10) is increasing with $r_0$.  At the same time,
for a fixed value of the throat radius $a$, the mass of the
thin shell is decreasing with $r_0$, as long as $r_0$ remains
much less than $a$.

\begin{figure}
\begin{center}
\vspace{0.5cm}
\includegraphics[width=0.5\textwidth]{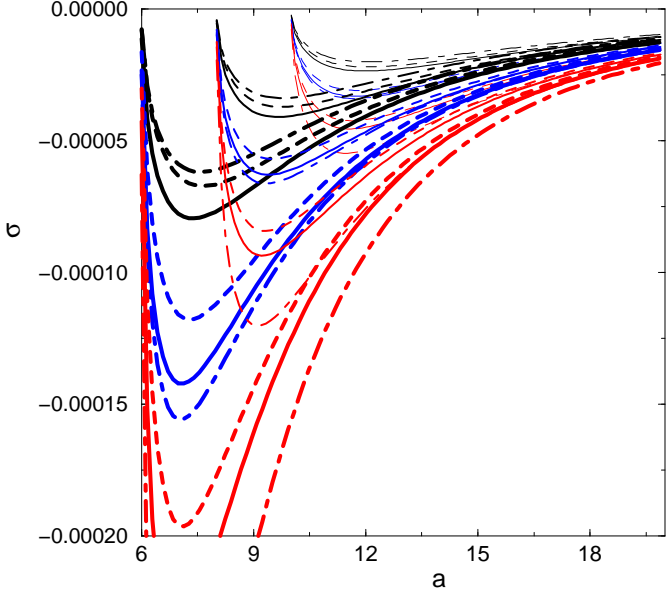}
\caption{Plot for $\sigma$ versus $a$. The description of the
curves is the same as in FIG.~\ref{fig1}. } \label{fig2}
\end{center}
\end{figure}



\section{Stability}
\noindent
Now we will focus our attention on the stability of the
configuration under small perturbations around a static solution
at $a_0$.  The starting point is the definition of a potential,
extended to our metric [Eq. (\ref{E:line1})].  Rearranging Eq.
(\ref{E:sigma1}), we obtain the thin shell's equation of
motion
\begin{equation}
\dot{a}^2 + V(a)= 0.
\end{equation}
Here the potential $V(a)$ is defined  as
\begin{equation}
V(a) =  f(a) - \left[\frac{4\pi
h(a)\sigma(a)}{h^{\prime}(a)}\right]^2.
\end{equation}
Expanding $V(a)$ around $a_0$, we obtain
\begin{eqnarray}
V(a) &=&  V(a_0) + V^\prime(a_0) ( a - a_0) +
\frac{1}{2} V^{\prime\prime}(a_0) ( a - a_0)^2  \nonumber \\
&\;& + O\left[( a - a_0)^3\right],
\end{eqnarray}
where the prime denotes the derivative with respect to $a$,
assuming a static solution situated at $a_0$. Since we are
linearizing around $ a = a_0 $, we must have $ V(a_0) = 0 $
and $V^\prime(a_0)= 0 $. The configuration will then be in
 stable equilibrium if $V^{\prime\prime}(a_0)>0$.

To carry out this analysis, we start with the energy
conservation equation.  Using Eqs. (\ref{E:sigma1}) and
(\ref{E:pressure1}), one can verify that
\begin{multline}\label{E:conservation}
  \frac{d}{d\tau}(\sigma\mathcal{A})+p\frac{d\mathcal{A}}
  {d\tau}=\\
    \{[h'(a)]^2-2h(a)h''(a)\}
   \frac{\dot{a}\sqrt{f(a)+\dot{a}^2}}{2h(a)},
\end{multline}
where $\mathcal{A}=4\pi h(a)$ by Eq.~(\ref{E:line1}).
The first term on the left side corresponds to a
change in the throat's internal energy, while the
second term corresponds to the work done by the
throat's internal forces.  According to
Ref. \cite{Eiroa2008a}, the right side
represents a flux.  From Eq. (\ref{E:conservation}), we get
\begin{multline*}
  \frac{d}{da}[\sigma h(a)]+\mathcal{P}
  \frac{d}{da}[h(a)]\\
   =-\{[h'(a)]^2-2h(a)h''(a)\}\frac{\sigma}{2h'(a)},
\end{multline*}
and, finally,
\begin{multline}\label{E:alternate}
  h(a)\sigma'+h'(a)(\sigma+p)+\{[h'(a)]^2-2h(a)
    h''(a)\}\frac{\sigma}{2h'(a)}\\=0.
\end{multline}
It is also shown in Ref. \cite{Eiroa2008a} that
\begin{multline}\label{E:Vdoubleprime}
  V''(a)=f''(a)+16\pi^2\times\\\left\{\left[\frac{h(a)}{h'(a)}
   \sigma'(a)+\left(1-\frac{h(a)h''(a)}{[h'(a)]^2}\right)
    \sigma(a)\right][\sigma(a)+2p(a)]\right.\\
   \left.+\frac{h(a)}{h'(a)}\sigma(a)[\sigma'(a)+2p'(a)]\right\}.
\end{multline}
Next, we define a parameter $\beta$, which is interpreted as the
subluminal sound speed, by the relation
\begin{equation}
\beta^2(\sigma) =\left. \frac{ \partial p}{\partial
\sigma}\right\vert_\sigma.
\end{equation}
To do so, observe that
\begin{multline*}
  \sigma'(a)+2p'(a)\\=\sigma'(a)[1+2p'(a)/\sigma'(a)]=
  \sigma'(a)(1+\beta^2).
\end{multline*}
Using Eq.~(\ref{E:alternate}),
we can now rewrite $V''(a)$ as follows:
\begin{multline}\label{E:potential}
 V''(a)=f''(a)-8\pi^2\left\{[\sigma(a)+2p(a)]^2
       \phantom{\frac{h}{h'}}\right.\\
   \left.+2\sigma(a)\left[\left(\frac{3}{2}-\frac{h(a)h''(a)}
   {[h'(a)]^2}\right)\sigma(a)+p(a)\right](1+2\beta^2)
   \right\}.
\end{multline}
At the static solution $a=a_0$, the conditions $V(a_0)=0$
and $V'(a_0)=0$ are indeed met.

Now consider the stability criterion $V''(a_0)>0$
starting with Eq. (\ref{E:potential}): first let
$V''(a_0)=0$ and solve for $\beta^2$.  We then find
that the graph of
 \begin{equation}\label{E:beta}
  \beta^2=-\frac{1}{2}+\frac{f''/8\pi^2-(\sigma+2p)^2}
   {4\sigma\left[\left(\frac{3}{2}-\frac{hh''}
     {(h')^2}\right)\sigma+p\right]}
\end{equation}
has a single vertical asymptote (Fig. 3.)  To the
right of the asymptote,
\begin{equation}\label{E:signchange}
   4\sigma\left[\left(\frac{3}{2}-\frac{hh''}
     {(h')^2}\right)\sigma+p\right]>0,
\end{equation}
also determined graphically.  Returning now to the
inequality $V''(a_0)>0$, we therefore have at $a=a_0$
\begin{equation}\label{E:above}
  \beta^2<-\frac{1}{2}+\frac{f''/8\pi^2-(\sigma+2p)^2}
   {4\sigma\left[\left(\frac{3}{2}-\frac{hh''}
     {(h')^2}\right)\sigma+p\right]}.
\end{equation}
So to the right of the asymptote, the region of
stability is below the graph of
Eq. (\ref{E:beta}), as shown in Fig. 3.

To the left of the asymptote, the sense of the
inequality in (\ref{E:signchange}) is reversed and
we obtain at $a=a_0$
\begin{equation}\label{E:below}
  \beta^2>-\frac{1}{2}+\frac{f''/8\pi^2-(\sigma+2p)^2}
   {4\sigma\left[\left(\frac{3}{2}-\frac{hh''}
     {(h')^2}\right)\sigma+p\right]}.
\end{equation}
So to the left of the asymptote, the region of stability
is above the graph.

\begin{figure}
\begin{center}
\vspace{0.5cm}
\includegraphics[width=0.5\textwidth]{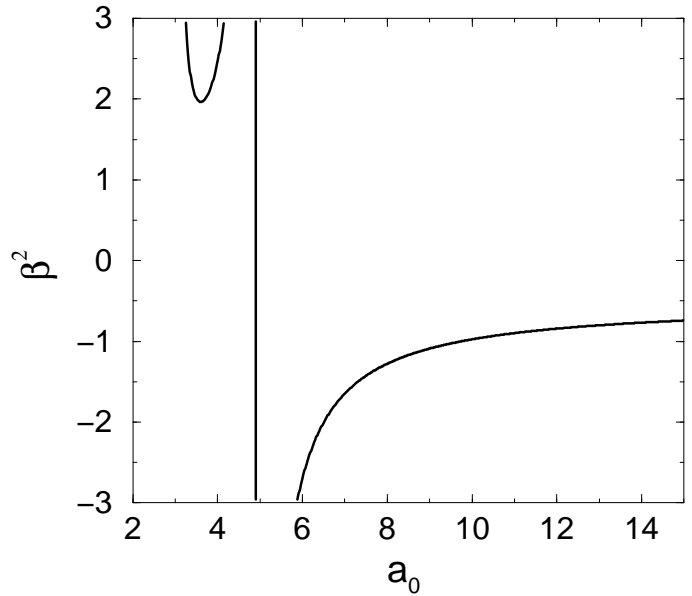}
\caption{Plot for $\beta^2$ versus $a_0$.
} \label{fig3}
\end{center}
\end{figure}

\begin{figure}
\begin{center}
\vspace{0.5cm}
\includegraphics[width=0.5\textwidth]{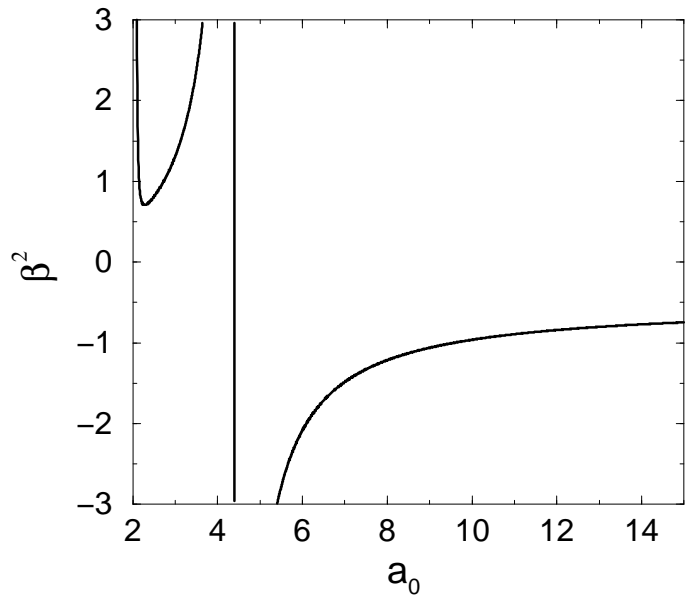}
\caption{All functions are the same. Some are cut off. See Fig.~3.
} \label{fig4}
\end{center}
\end{figure}

Fig.~3 does indeed show typical regions of stability
using somewhat arbitrary values of the various
parameters: $r_0=1$, $r_-=2$, $r_+=3$, and $n=1/2$.
As noted above, the region is below the curve on the
right and above the curve on the left.  The sign change
is determined by inequality (\ref{E:signchange}).
These results, including the graphs, are similar to
those in Refs. \cite{Poisson1995} and \cite{Eiroa2008a}
(dealing with Schwarzschild and dilaton thin-shell
wormholes, respectively) in the sense that the regions
do not correspond to any value in the interval
$0<\beta^2\le 1$.  Since $\beta$ is ordinarily
interpreted as the speed of sound, it is highly desirable
to obtain a region for which $\beta^2<1$.  This is
indeed possible for our wormhole: if we choose $r=r_-$
close to $r=r_+$, then we typically get a region of
stability for $\beta^2<1$.  For example, in Fig. 4,
$r_0=1$, $r_-=2$, $r_+=2.05$, and $n=0.8$.  The closer
$r_-$ is to $r_+$, the more the region of stability
extends below $\beta^2=1$.
\\
\\
\\

\section{Conclusion}\noindent
A new black-hole solution by Sur, \emph{et al.}, for
Einstein-Maxwell scalar fields was inspired by low-energy
string theory.  This paper discusses a new thin-shell
wormhole constructed by applying the cut-and-paste
technique to two copies of such black holes.  We analyzed
various aspects of this wormhole, such as the amount of
exotic matter required, the attractive or repulsive
nature of the wormhole, and a possible equation of state
for the thin shell.  The stability analysis concentrated
on the parameter $\beta$, normally interpreted as the
speed of sound.  It was found that whenever the two
event horizons are close together, a stability
region exists for some values of $\beta^2$ less
than unity, unlike the the cases discussed in
Refs. \cite{Poisson1995} and \cite{Eiroa2008a}
for Schwarzschild and dilaton thin-shell
wormholes, respectively.

\subsection*{Acknowledgments}

AAU, FR and SR are thankful to Inter-University Centre for
Astronomy and Astrophysics, Pune, India for providing Visiting
Associateship under which a part of this work is carried out. FR
and ZH are also grateful to UGC, Govt. of India and D.S. Kothari
fellowship,
 for financial
support.

\end{document}